\documentclass[aps,preprint,eqsecnum,amssymb]{revtex4}
\usepackage{graphics}
\usepackage{bm}
\newcommand{\be}{\begin{equation}}
\newcommand{\ee}{\end{equation}}
\newcommand{\bea}{\begin{eqnarray}}
\newcommand{\eea}{\end{eqnarray}}
\newcommand{\pt}{\partial}
\newcommand{\tr}{\mbox{ Tr} }
\newcommand{\ra}{\rightarrow}
\begin{document}
\title{ Quark-Antiquark Potential for Siegel Strings}
\author{ M. Arshad Momen and G. Dastegir Al-Quaderi}\thanks{Current Address: St. Edmund's College, 
Univ. of Cambridge, CB3 0BN, UK}
\affiliation{Department of Physics, Dhaka University, Dhaka-1000, Bangladesh}
\begin{abstract}
We compute the quark-antiquark potential employing the continuum
action for QCD-like random lattice strings proposed by Siegel. The model leads to 
a potential similar to those obtained from Nambu-Goto string theory but has 
some modifications which we interpret as velocity dependent contributions. We 
also propose to add extra terms in the action which lead to physically interesting
propagators for partons for the infrared region. 
\end{abstract}
\maketitle

\section{Introduction}

Despite the currently established status of QCD as {\em the} theory for strong interactions,  
due to its non-perturbative nature in the infrared regime,
it is desirable to seek a simpler description of low-energy hadron 
spectrum in terms of phenomenological models. Often these models involve some type of 
non-trivial classical structure like bags, strings or instantons. Amogn these strings were  
historically introduced to explain the so-called Regge behavior and duality 
symmetry demonstrated by the hadron-hadron scattering amplitudes at high energies. 
However, quantization of the strings as fundamental {\em hadronic } 
objects suffered from several severe drawbacks: the presence of unphysical particles in the 
spectrum ( tachyons, massless spin-2 particles etc. ) as well as requiring a 26-dimensional 
spacetime. The subsequent advent of the QCD has since then pushed the hadronic string model 
into its current status as a phenomenological attempt to explain the hadronic scattering. 
Fundamental strings have instead taken up the role of the prime model for the unification of 
fundamental interactions. 
 
\begin{figure}
\includegraphics{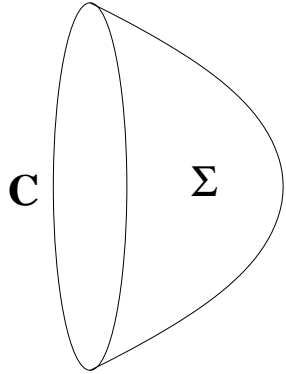}
\caption{ String worldsheet $\Sigma$ bounded by the loop $C$ over which supports the 
Wilson loop}
\label{fig1}
\end{figure}

Despite the false start, string theory received some theoretical justification from QCD
when  't Hooft  \cite{thooft}  showed that  $SU(N)$ gauge theories behaved as 
an effective planar ( hence a two-dimensional ) theory in the limit $ N \rightarrow \infty$,
provided  color is  confined. An earlier connection between string theory and large gauge
theories was provided by the Nambu conjecture relating the Wilson loop operator
( defined in gauge theoretic context) defined over the closed loop C to the string 
theory defined on the world sheet $\Sigma$ bounded  $C$ ( cf. Fig \ref{fig1} ):
\be
W[C] \sim e^{-S[\Sigma]},
\label{0.1}
\ee
Here the string is assumed to be moving freely in a flat spacetime. 
In the recent years, this correspondence has been put on firmer foundations   
by Maldacena\cite{maldacena} and Witten \cite{witten} which currently goes by the 
name of AdS/CFT correspondence \cite{gubser} where the low-energy 
string theory is represented by five dimensional AdS supergravity theory and the 
corresponding four-dimensional gauge theory lives on the  minkowskian ( conformal )
boundary of the AdS spacetime. 

An alternative path, explored by Nielsen and others \cite{sakita}, is to show  the
amplitudes of the dual models can be reproduced from scalar field theories  ( representing the
``partons'' ) defined on two-dimensional spacetime using Gaussian propagators for the scalar fields. 
Though such propagators provide the field theoretic reason behind the better high-energy 
behavior of the scattering amplitudes of strings, the underlying field theory needs to be 
a non-polynomial theory. Also, it is known that Gaussian propagators behave as 
massive propagators for low value of momentum,
$\Lambda^2 e^{-\frac{p^2} {\Lambda^2}}\sim 
\frac{1}{\Lambda^2 + p^2}$ which contradicts our knowledge of 
perturbative QCD. This tells us that such ``partonic'' models do not corresponds to 
QCD string.

Therefore, it is natural to attempt to construct a string action 
action whose ``partons'' are massless. Indeed,sometime ago Siegel has derived 
such an action \cite{siegel}. This action is particularly interesting from the fact that 
it correctly predicts the spacetime dimension to be four by demanding T-duality invariance of the 
 string action.  However, to explain the asymptotic freedom one has to assume that the
interactions in this theory be described by the wrong sign $\phi^4$ theory. Though, this 
theory will suffer from the problems with vacuum instabilities we will not concern ourselves here.

An important feature that emerges from  string models is that
 a linearly confining  potential is obtained in the leading order from string actions 
proportional to the worldsheet area :
\be
V(R) = \tau R  + \cdots
\label{0.2}
\ee
where $\tau $ is the string tension and the sub-leading order terms 
are represented by the ellipsis.

In this brief report, we attempt to check whether this theory manages to reproduce 
the typical potential between quark-antiquark calculated from string theories. Unlike 
the Nambu or Polyakov strings the Siegel string involves a traceless symmetric matrix 
apart from the regular worldsheet metric. We find that the off-diagonal 
elements of this matrix contributes to the one-loop effective potential. We 
suggest that these off-diagonal elements are related to the velocity 
of the endpoints of the strings where the quarks reside. However, the matrix elements
are not fixed in this theory. This problem is quite similar to the runaway problem of 
the moduli fields in fundamental string theory. However, unlike the fundamental string 
theory writing a potential for this matrix field is not prohibited anything like 
general coordinate invariance or supersymmetry. We argue that a realistic 
potential modifies the partonic propagators in the infrared regime which 
is reasonable for a string picture. 
 
The paper is arranged as follows. We first sketch Siegel's construction of the
action for QCD-like strings in section 2. In  section 3, we carry out the one-loop computation 
of the quark-antiquark potential using Siegel's action following the procedures of L\"{u}scher
et. al. \cite{luscher} and  Alvarez
\cite{alvarez}. We contrast our results with the well-known results \cite{luscher} in the 
literature. Finally, we comment on our results and on its physical implications.
  
\section{Siegel's action for QCD-like strings}

Siegel's construction of the action stems from the old observation due
to Nielsen and others  \cite{sakita} that one can reproduce the amplitudes of 
the dual models by working with a two dimensional scalar field theory whose propagators are 
Gaussian in the momentum space. Therefore, it is interesting to search for a string theory which 
``originates'' from a theory of massless propagators. Following the random triangulation approach 
for low-dimensional strings, we consider the worldsheet to be represented 
by a random lattice. The continuum limit will be recovered when the number of points on the 
lattice $N \ra \infty$.   

\begin{figure}
\includegraphics{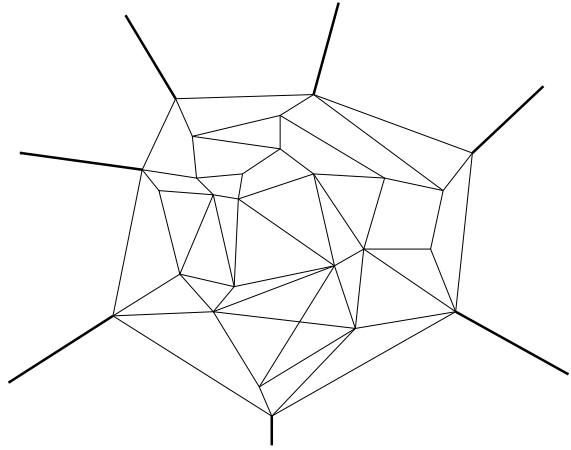}
\caption{ A random lattice graph which makes up the string worldsheet}
\label{fig2}
\end{figure}

One begins with the Fock-Schwinger-DeWitt propertime parameterization of the massless propagators, 
namely
\be
\frac{1}{p^2} = \int^\infty_0 d \tau e^{- \tau p^2}.
\label{1.1}
\ee
 Consider a Feynman graph involving massless scalar fields on a random lattice ( cf. Fig 
\ref{fig2} )  whose points are 
labelled by integers $i=1, 2, \cdots, N$ with $N \rightarrow
 \infty$. The propagators denoted by $p_{ij}$ , being two point functions,  are 
 elements of a $N \times N$ matrix and supported on the link connecting the lattice points $i$ 
and $j$ . The amplitude in momentum space associated with this graph will be given by 
\be
{\cal A} \sim \int \prod_{<ij>} d^D p_{ij} \frac{1}{p^2_{ij}} \prod_{i} \delta ( \sum_j p_{ij})
\label{1.2}
\ee
where the delta function simply enforce the momentum conservation at all vertices.  Using the 
Schwinger parameterization (\ref{1.1}) in (\ref{1.2}) we rewrite the amplitude as
\be
{\cal A} \sim \int \prod_{<ij>} d^D p_{ij} d \tau_{ij} e^{- \tau_{ij} p^2_{ij}}\int  \prod_i dX_i
 e^{i \sum_{j} p_{ij} X_i}
\label{1.3}
\ee
where we have used a matrix-valued Schwinger parameter. The momentum conservation equation  
at the vertices can be solved by going to the loop representation for the graph, i.e. by writing
\be
p_{ij} \equiv p_{IJ} =  k_I - k_J
\label{1.3.1}
\ee
where $k_I, k_J$ are the loop momenta running over the $I$th and $J$th loop, whose intersection 
is the link represented by the pair $ij$.
momenta to obtain 
\be
{\cal A} \sim \int  \prod_I d \tau_{IJ} e^{ - \tau_{IJ} ( k_I- k_J)^2}
\label{1.3.2}
\ee
where we have switched to the loop representation. 

To obtain the continuum limit of this action one has to note that there are only two independent 
directions on the string worldsheet one demands that the matrix has two degrees of freedom - so 
this matrix must be a traceless symmetric matrix. 

The final result is the  Siegel's action is of the form 
\be
S_0 = \int d^2 \sigma \left[ \tau^{mn} \pt_m X \cdot \pt_n X + \lambda g^{mn} \tau_{mn} + 
\cdots\right]
\label{1.4}
\ee
where $g_{mn}$ is the worldsheet metric used for taking the trace of $\tau$ and the last term 
imposes the tracelessness condition via the Lagrange multiplier $\lambda$.

\section{Calculation of the quark-antiquark potential}

The calculation of the quark-antiquark potential using the string models was pioneered by 
L\"{u}scher, Symanzik and Weisz \cite{luscher} using the Nambu-Goto action. Further improvements 
of this result were done by Alvarez\cite{alvarez} and Arvis \cite{arvis}. 

\begin{figure}
\includegraphics{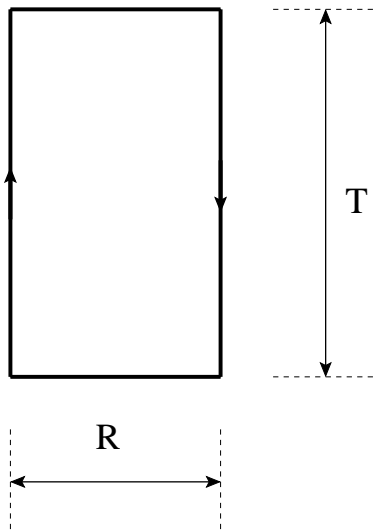}
\caption{ Quark-Antiquark Loop used for calculating the Quark-antiquark Potential}
\label{fig3}
\end{figure}

The calculation proceeds via considering a quark-antiquark virtual rectangular loop where the 
separation between the quark-antiquark pair is $R$ and the pair propagates for time $T$ as in 
figure \ref{fig3}.
In the confining phase, the Wilson criterion gives
\be
< W(C)> \sim e^{- T V(R)} 
\label{2.0}
\ee  
Using the Nambu conjecture one can relate the string action with the quark-antiquark potential:
\be
V(R)= \frac{1}{T} S(\Sigma) 
\label{2.0.5}
\ee
Thus in flat spacetimes when the string action is proportional to the worldsheet ( hence the loop)
area one recovers a linear potential in the tree level. However, at short distances the quantum 
fluctuations can have important consequences for instance determining the deconfinement behavior 
\cite{pisarski}. Therefore it is important to calculate the one-loop contribution to the 
quark-antiquark potential.

We begin our calculations by embedding the worldsheet in the spacetime as
\be
\tau = t = X^0, \qquad \qquad \sigma=X^1
\label{2.2}
\ee
The rest of the coordinates act as transverse excitations on the worldsheet and contribute to the 
effective potential via quantum fluctuations.      

Let us parameterize the matrix $\tau_{mn}$ as follows
\be
\tau = \left( 
\begin{array}{cc}
	a & b \\
	b & a 
\end{array}
\right)
\label{2.3}
\ee
which is traceless with respect to a Minkowskian metric. As briefly mentioned earlier, 
the expectation values of the fields $a, b$ are not fixed due to the absence of any classical 
potential for them. However, following the  $\sigma$-model approach to string theory 
\cite{fradkin} we assign a {\em constant} vacuum expectation value for $a,b$ ( this type of 
approximation was also used by Alvarez \cite{alvarez} ). 

Using the embedding (\ref{2.2}) the action now reads 
\be
S= 2 a T R + \int d^2 \sigma \tau^{mn} \pt_m \phi \cdot \pt_n \phi + \cdots
\label{2.4}
\ee
One can see that the diagonal matrix element $a$  is gives rise to the tree level string 
tension. However, the transverse fluctuations also contribute to the string potential which 
is to be interpreted as the contribution from the zero point fluctuations about the classical 
configuration (\ref{2.2}) ( Casimir energy ). If we integrate out the transverse fields 
the ``effective action'' reads
\be
S= 2a T R + \frac{(D-2)}{2} \tr \ln \left[ \pt_m \tau^{mn} \pt_n \right]   
\label{2.5}
\ee
 
To find a real effective action we have to guarantee that the eigenvalues of the operator
appearing in the second term of (\ref{2.5}) be real - hence the operator be self-adjoint.
This can be done by choosing appropriate boundary conditions \cite{richtmyer} 
as the worldsheet has boundaries.
The self-adjointness condition for the operator appearing in (\ref{2.5}) would read as 
\be
( \chi, {\cal D} \psi ) = ( {\cal D} \chi , \psi ) \qquad \forall \psi,\chi \in M
\label{2.6}
\ee
where $M$ is the domain of the operator ${\cal D}$\cite{richtmyer}:
\be
{\cal D} \equiv \pt_m ( \tau^{mn}\pt_n) 
\label{2.7}
\ee
specified by the chosen boundary condition. In equation (\ref{2.6}) the inner product is defined 
is the standard fashion, namely,
\be
(\eta, \xi) \equiv \int_0^R d \sigma\, \eta^\dagger(\sigma) \,\xi(\sigma)
\label{2.7.1}
\ee 

We are interested in the stationary configurations of the string and accordingly seek
stationary solutions of the form 
\be
\psi(\sigma, t) = \psi(\sigma) e^{-i \omega t}
\label{2.7.1.5}
\ee
Restricted to this type of subspace, the operator ${\cal D}$ takes the form 
\be
{\cal D}_\omega = \left[ a ( \pt_\sigma^2 - \omega^2 ) - 2 i b \omega \pt_\sigma \right]
\label{2.7.1.6}
\ee
   
Due to the presence of the worldsheet boundaries, the condition (\ref{2.6}) for the operator
${\cal D}_\omega $  now translates into the vanishing of the following anti-hermitian 
bilinear form
\be
\bm{\zeta}^\dagger {\cal A} \bm{\chi} = 0
\label{2.7.2}
\ee
where the vectors $\bm{\chi}, \bm{\zeta}$ are of the form
\be
\bm{\zeta} \equiv \left( \begin{array}{c} \psi(0) \\ \frac{1}{\omega}\pt_{\sigma} \psi (0) \\ 
\psi(R) \\ \frac{1}{\omega}\pt_{\sigma} \psi(R) \end{array}
\right) 
\label{2.7.3}
\ee
and 
\be
{\cal A} \equiv \left[ \begin{array}{cccc} 2 i \tan \alpha  & ~1 & ~0 & ~0\\
-1 & ~0 & ~0 & ~0 \\
~0 & ~0 &  -2 i \tan \alpha & -1 \\
~0 & ~0 & ~1 & ~0 
\end{array} \right]
\label{2.7.4}
\ee
where $\tan \alpha \equiv \frac{b}{a} \in (-\frac{\pi}{4}, \frac{\pi}{4})$. The limits follow
from the conditions that the eigenvalues of the matrix $\tau$ must be positive and accordingly 
$\mbox{det }\tau = ( a^2 - b^2) > 0$. Therefore $ \frac{b}{a}^2 < 1$.
  
The eigenvalues $\mu_i$ and the corresponding eigenvectors $\Psi_i$ with $ i=1 \cdots 4$ 
of the matrix ${\cal A}$ are as follows:
\be
\mu_1 = i  \tan (\frac{\alpha}{2} + \frac{\pi}{4}) ) \qquad \qquad \Psi_1 = \left( 
\begin{array}{c} \cos (\frac{\alpha}{2} - \frac{\pi}{4})  \\
i \sin( \frac{\alpha}{2} - \frac{\pi}{4}) \\
0 \\
0
\end{array},
\right)
\label{2.7.5}
\ee
\be
\mu_2 = i  \tan(\frac{\alpha}{2} - \frac{\pi}{4} ) \qquad \qquad \Psi_2 = \left( 
\begin{array}{c} \cos (\frac{\alpha}{2} + \frac{\pi}{4})  \\ \sin 
(\frac{\alpha}{2} + \frac{\pi}{4})\\
0 \\
0
\end{array},
\right)
\label{2.7.6}
\ee
\be
\mu_3 = -i \tan (\frac{\alpha}{2} + \frac{\pi}{4})\\
\qquad \qquad \Psi_3 =  \left( 
\begin{array}{c} 0\\0 \\
 \cos( \frac{\alpha}{4} -  \frac{\pi}{4}) \\
i \sin (\frac{\alpha}{4} -  \frac{\pi}{4})
\end{array}
\right)
\label{2.7.7}
\ee
\be
\mu_4 = -i \tan ( \frac{\alpha}{2}- \frac{\pi}{4}) \qquad \qquad \Psi_4 = \left(
\begin{array}{c} 0\\
0\\
\cos (\frac{\alpha}{2 } +  \frac{\pi}{4})\\
i \sin( \frac{\alpha}{2} +  \frac{\pi}{4}) 
\end{array},
\right)
\label{2.7.8}
\ee
the eigenvectors have been normalized here.

Domains over which the operator ${\cal D}_\omega $ is self-adjoint can now be found by 
taking various combinations of the eigenvectors (\ref{2.7.5},\ref{2.7.6},\ref{2.7.7},\ref{2.7.8})
with relative U(1) phase rotations between them.Therefore, any function$ \bm{\chi}$ belonging to the domain $M$ 
 will be given by a general linear combination of the form
\be
\bm{\chi} = \Psi_1 + e^{i \theta} \Psi_2 + e^{i \phi}\left( \Psi_3 + e^{i \xi} \Psi_4 \right)
\label{2.7.9}
\ee
where $\theta,\phi,\xi$ are three $U(1)$ phases. 
They parameterize the various boundary conditions compatible with
self-adjointness of ${ \cal D}_\omega$.

Written explicitly (\ref{2.7.9}) reads,
\be
\left(
\begin{array}{c}
\psi(0)\\
\frac{1}{\omega}\pt_\sigma \psi(0) \\
\psi(R)\\
\frac{1}{\omega}\pt_\sigma \psi(R)
\end{array} \right) = 
\left( \begin{array}{c} 
\cos ( \frac{\alpha}{2} - \frac{\pi}{4} ) + e^{i\theta} \cos ( \frac{\alpha}{2} + 
\frac{\pi}{4})\\
-i (\sin ( \frac{\alpha}{2} - \frac{\pi}{4} ) + e^{i\theta} \sin ( \frac{\alpha}{2} + 
\frac{\pi}{4}
))\\ 
e^{i \phi} ( \cos ( \frac{\alpha}{2} - \frac{\pi}{4} ) + e^{i\xi} \cos ( 
\frac{\alpha}{2} + \frac{\pi}{4})) \\
-i e^{i \phi} ( \sin ( \frac{\alpha}{2} - \frac{\pi}{4} ) + e^{i\xi} \sin ( 
\frac{\alpha}{2} + \frac{\pi}{4}))
\end{array}
\right)
\label{2.8.0}
\ee
   
Note that the boundary values of $\psi$ at the point $\sigma=0$ gets contribution from 
the eigenvectors $\Psi_1$ and $\Psi_2$ while the boundary conditions at $\sigma = R$ 
gets contribution from $\Psi_3$ and $\Psi_4$.

The eigenvalues of the operator ${\cal D}_\omega$  are then found by solving by the equation
\be
a \left[( \pt_{\sigma}^2 - \omega^2 ) -2 i  \omega \tan \alpha ~\pt_{\sigma} \right] \psi( \sigma) = \lambda \psi(\sigma)
\label{2.8.3}
\ee
subject to the boundary conditions (\ref{2.8.1},\ref{2.8.2}). We seek solutions of the form 
\be
\psi ( \sigma ) \sim e^{i p \sigma }
\label{2.8.4}
\ee
Correspondingly, one finds the general solution is a given by
\be
\psi ( \sigma)  = A e^{i p_+ \sigma } + B e^{i p_- \sigma}
\label{2.8.5}
\ee
with 
\be
p_\pm = \omega \tan \alpha \pm \sqrt{ \omega^2  \sec^2 \alpha  + \frac{\lambda}{a}}
\label{2.8.6}
\ee

Using (\ref{2.8.6}) in the equation ( \ref{2.8.0}) we are led to a set of equations which are 
underdetermined. This situation can be remedied by making the choice $\theta=\xi$ which allows 
us to determine the eigenvalues $\lambda$ readily : 
\be
\lambda_n= ( a- \frac{b^2}{a}) \omega^2 + a \left( \frac{n \pi}{R}\right)^2, 
\label{2.9}
\ee
where $ n $ is an integer. 

Note that when $\theta= \xi$, the boundary conditions lead to the phase relation
\be
\psi(0) = e^{i \phi} \psi(R) \qquad \pt_\sigma \psi(0) = e^{i \phi} \pt_\sigma \psi(R) 
\label{2.8.1}
\ee
It is therefore very suggestive to interpret the angle $\phi$ as a Aharonov-Bohm phase 
factor between the end points of the strings, where the quarks are placed. With the hindsights 
obtained from QCD, one can suggest that this phase arises due to chromoelectric field stretched 
between the quarks at the endpoints. However as it is impossible to calculate this phase within
this model this can only arise in terms of as an self-adjoint extension parameter entering 
at the boundaries.    

Also within this restricted set of boundary conditions one finds also that at the end themselves:
\be
\left.\left({ \cos \alpha +e^{i \theta} ( 1+ \sin \alpha )} \pt_{\sigma} \psi\right) +i 
 \omega \left({e^{i \theta} \cos \alpha + ( 1+ \sin \alpha )} \right) \psi \right|_{0}^{R} = 0
\label{2.8.2}
\ee
Mixed boundary conditions of this type has appeared in the string literature in the discussion of 
Neumann-Dirichlet tadpoles \cite{kogan}.  If $\psi$ is to be interpreted as a single particle 
wavefunction then one can relate the ratio $\frac{b}{a}$ to  the transverse velocity 
of the string ends with respect to the string center of mass. 

Carrying out the functional sum over all the eigenvalues, following \cite{alvarez}  one is 
immediately led to the following potential 
\be
V(R) = 2 a R - \frac{(D-2) \pi}{12 R} \frac{1}{\sqrt{1 - (\frac{b}{a})^2}}
\label{2.10}
\ee
Note that the potential again depends on the ratio $\frac{b}{a}$. When $b$ is set equal to 
zero we recover the original result due to L\"{u}scher et. al. \cite{luscher}. Note that 
if we hold  $det \tau$ fixed then one can think of the tuple $(a,b)$ as a 2d vector on a  
minkowskian worldsheet whose time component is $a$ and the spatial component is
( directed along the length of the string)  $b$. Thus the ``spatial velocity'' would
be given by the ratio $\frac{b}{a}$. The quantum correction to the classical potential
then has the correct ``Lorentz transformation'' factor. 
Note also that the model doesn't require any condition on  spacetime dimension $D$ as 
long as $D \geq 2$, unlike the treatment in  \cite{alvarez} where one requires $D \ra \infty$.  
This features makes the Siegel's  string  a more general model for hadronic 
strings compared to the Nambu-Goto string.  

However, from equation (\ref{2.10}) it appears that the parameters $a, b$ are  arbitrary as long 
as the trace and determinant of $\tau$remains positive, i.e. $ a > 0$ and $ a^2 > b^2 $. To 
stabilize the value of $\frac{b}{a}$ one can introduce potentials for the matrix $\tau_{mn}$ at 
the tree level.  For instance, if we restrict ourselves to quadratic and higher order 
potentials we can write
\be
V( \tau)  = g_1 \tr ( \tau^2) + g_2 det ( \tau) + \cdots 
\label{2.11}
\ee
where $g_1, ~g_2$ are small coupling constants and the ellipsis represents higher order polynomials
of $\tau$. A short calculation shows that a minima of the potential is obtained for fixed $R$ if 
$g_1$ is negative ( provided $a > 0$, so that the string tension is positive ).
 
In presence of such potentials for $\tau$ the partonic propagators are modified in the 
infrared regime, specifically picking up higher order poles at $p=0$. For instance, if we use the 
quadratic potential in (\ref{2.11}), the Fock-Schwinger-Dewitt propertime 
parameterization lead to the following propagator ( symbolically) 
\bea
\int && d \tau e^{- (\tau p^2 + g \tau^2 + \cdots)} \sim \int d \tau e^{ -\tau p^2} ( 1 + 
g  \tau^2 + \cdots) \\
&& = \frac{1}{p^2} + \frac{\lambda}{2 p^4} + \cdots
\label{2.12}
\eea
where $g$ represents both the coupling constants in (\ref{2.11}).

The appearance of the $\frac{1}{p^4}$ is interesting from the fact that such singular 
propagators lead to confining potentials. Singular propagators of this type also appear in 
the dual superconductor picture for QCD  \cite{ball} described by  Higgs-type models 
which curiously supports stringy Nielsen-Olesen vortices\cite{NO}. This suggests that if one uses propagators with 
higher order infrared singularities for the partons,  the corresponding string 
theory might contain potential terms for the ``moduli'' field $\tau$.

\section{Comments}

We have examined the Siegel's action for QCD-like strings as an viable model by calculating the  
quark-antiquark potential in this model which is more general than the results of L\"{u}scher 
et. al. As stated earlier, in this model the matrix field $\tau_{mn}$ suffers from the same type 
of runaway  problem as the moduli fields in fundamental strings.
However, unlike moduli fields whose potential are prohibited by general covariance or 
supersymmetry one can stabilize the field $\tau_{mn}$ by introducing a potential for this model. 
We argued that the role of such potentials is to modify the partonic propagators at low-energies. 
However, such behavior could be welcomed from the view of dual-superconductor picture for QCD 
which sustains composite stringy solutions\cite{ball} like the Abrikosov flux tubes for Type II 
superconductors.

\end{document}